\documentclass[twocolumn]{aastex631}
\usepackage{CJK}

\pdfoutput=1
\newcommand{\mesaone}{Paper~I}

\shorttitle{The statistical analysis of Li-rich and normal giants}
\shortauthors{Zhou et al.}

\begin{document}
\begin{CJK*}{UTF8}{gbsn}
\title{Li-rich Giants in LAMOST Survey. III. The statistical analysis of Li-rich giants}

\author[0000-0002-4391-2822]{Yutao Zhou}
\altaffiliation{LAMOST FELLOW}
\affiliation{Department of Astronomy, School of Physics, Peking University, Beijing 100871, People's Republic of China}
\affiliation{Kavli Institute for Astronomy and Astrophysics, Peking University, Beijing 100871, People's Republic of China}

\author{Chun Wang}
\affiliation{Tianjin Astrophysics Center, Tianjin Normal University, Tianjin 300387, People's Republic of China.}

\author[0000-0002-8609-3599]{Hongliang Yan}
\affiliation{Key Laboratory of Optical Astronomy, National Astronomical Observatories, Chinese Academy of Sciences, Beijing 100101, People's Republic of China}
\affiliation{University of Chinese Academy of Sciences, Beijing 100049, People's Republic of China}

\author[0000-0003-3250-2876]{Yang Huang}
\affiliation{South-Western Institute for Astronomy Research, Yunnan University, Kunming 650500, People's Republic of China}

\author[0000-0002-6434-7201]{Bo Zhang}
\affiliation{Department of Astronomy, Beijing Normal University, Beijing 100875, People's Republic of China}

\author[0000-0001-5082-9536]{Yuan-Sen Ting (丁源森)}
\affiliation{Research School of Astronomy $\&$ Astrophysics, Mount Stromlo Observatory, Cotter Road, Weston Creek, ACT 2611, Canberra, Australia}
\affiliation{Research School of Computer Science, Australian National University, Acton ACT 2601, Australia}

\author[0000-0002-7727-1699]{Huawei Zhang}
\affiliation{Department of Astronomy, School of Physics, Peking University, Beijing 100871, People's Republic of China}
\affiliation{Kavli Institute for Astronomy and Astrophysics, Peking University, Beijing 100871, People's Republic of China}

\author[0000-0002-0349-7839]{Jianrong Shi}
\affiliation{Key Laboratory of Optical Astronomy, National Astronomical Observatories, Chinese Academy of Sciences, Beijing 100101, People's Republic of China}
\affiliation{University of Chinese Academy of Sciences, Beijing 100049, People's Republic of China}

\altaffiliation{LAMOST FELLOWSHIP}
\correspondingauthor{Huawei Zhang, Jianrong Shi}
\email{zhanghw@pku.edu.cn, sjr@nao.cas.cn}

\begin{abstract}
The puzzle of Li-rich giant is still unsolved, contradicting the prediction of the standard stellar models. Although the exact evolutionary stages play a key role in the knowledge of Li-rich giants, a limited number of Li-rich giants have been taken with high-quality asteroseismic parameters to clearly distinguish the stellar evolutionary stages. Based on the LAMOST Data Release 7 (DR7), we applied a data-driven neural network method to derive the parameters for giant stars, which contain the largest number of Li-rich giants. The red giant stars are classified into three stages of Red Giant Branch (RGB), Primary Red Clump (PRC), and Secondary Red Clump (SRC) relying on the estimated asteroseismic parameters. In the statistical analysis of the properties (i.e. stellar mass, carbon, nitrogen, Li-rich distribution, and frequency) of Li-rich giants, we found that: (1) Most of the Li-rich RGB stars are suggested to be the descendants of Li-rich pre-RGB stars and/or the result of engulfment of planet or substellar companions; (2) The massive Li-rich SRC stars could be the natural consequence of Li depletion from the high-mass Li-rich RGB stars. (3) Internal mixing processes near the helium flash can account for the phenomenon of Li-rich on PRC that dominated the Li-rich giants. Based on the comparison of [C/N] distributions between Li-rich and normal PRC stars, the Li-enriched processes probably depend on the stellar mass.

\end{abstract}

\section{Introduction}  \label{sec:sec1}
\end{CJK*}
Lithium is one of the most important elements produced at the early universe \citep{2021Randich}. It is easily burnt at the temperature of several million Kelvin. As a low mass star evolves out of the main-sequence (MS), the Li material will be transported from the stellar surface to the interior by the first dredge-up (FDU) process, which leads to the depletion of surface Li \citep{1967Ibena, 1967Ibenb}. The stellar evolution theory suggests that Li will be further depleted when the star climbs up to the red giant branch (RGB) bump, where the extra mixing conveys more Li to hot regions \citep{1994Charbonnel, 1995Charbonnel}. Against the expectation of these models, a few percent of giants, namely Li-rich giants, present Li abundances A(Li) greater than 1.5 dex \citep{1982Wallerstein, 1989Brown, 2011Kumar, 2016Casey, 2018Zhou, 2018Yan, 2019Gao, 2021Deepak, 2021Martell}.

After the first Li-rich giant discovered by \citet{1982Wallerstein}, researchers have paid much attention to such object \citep{2000Charbonnel, 2002Drake, 2011Kumar, 2013Martell, 2016Casey, 2018Smiljanic, 2019Zhou}. However, the sample size is limited. Currently, large scale surveys have been available for studying these rare stars, such as the GALactic Archaeology with HERMES (GALAH, \citealp{2019Deepak, 2021Martell}), the Large Sky Area Multi-Object Fiber Spectroscopic Telescope (LAMOST, \citealp{2019Gao, 2019Casey}), and the number of Li-rich giants has reached about ten thousand. The consensus of origin of Li production is not reached. Other than the extra mixing to destroy Li, the processes of ``enhanced" extra mixing with a high mixing rate are proposed to be responsible for the Li enrichment, which can produce more Li because of high mixing efficiency. But the exact physical mechanism to trigger the mixing is still uncertain. The theories are mainly the thermohaline mixing \citep{2006Eggleton, 2007Charbonnel}, rotational mixing \citep{2004Denissenkov, 2019Casey} and magneto-induced mixing \citep{2007Busso}.

Many explanations of Li enhancement are connected to the special evolutionary stages and the exact evolutionary stage of a star is hard to tell due to the overlap of different evolutionary stages, especially between the RGB bump and the red clump (RC). In the past few decades, the Li-rich giants are supposed to be the RGB bump within a narrow region of Hertzsprung$-$Russell (HR) diagram \citep{2000Charbonnel}. When the hydrogen-burning shell burns out the barrier of the mean molecular weight left after the FDU, it enables the extra mixing such as the thermohaline instability \citep{2010Charbonnel}. It is reasonable to consider the Li-rich giants at the RGB bump because of the long time-scale in this stage associated with the extra-mixing. The first Li-rich giant with helium-core burning is reported based on the asteroseismic analysis by \citet{2014Silva}. As \citet{2011Bedding} suggested, the RC and RGB stars can be well separated by the asteroseismic parameters, especially the large frequency separation ($\Delta \nu$) and the period spacing ($\Delta P$). Recently, \citet{2021Yan} provided a uniquely large sample with a clean determination of evolutionary stages based on the asteroseismology, they found that most ($\sim 80\%$) of the Li-rich giants belong to the phase of RC rather than RGB. A similar result has also presented by \citet{2021Singh}.

Although the evolutionary stages can be well distinguished by asteroseismology, the number of Li-rich giants with asteroseismic information is limited. More Li-rich RC stars are identified by the data-driven method though they are not directly determined from the asteroseismic data \citep{2019Casey, 2019Zhou, 2020Kumar}. Based on the location on the HR diagram associated with data-driven method, \citet{2020Kumar} found that the observed Li in the RC stars (A(Li) $\sim$ 0.7) are enhanced by about 1.6 dex relative to the prediction of models. They claimed that all the low-mass RC stars have gone through a process of Li production. This process is suggested to operate during the helium flash or the RGB tip \citep{2020Schwab,2021Mori,2021Zhangjh}, which is against by \citet{2021Chaname}. \citet{2021Chaname} argued that this finding is misled by stellar masses because the masses estimated by \citet{2020Kumar} are higher than the reliable asteroseismic masses, which cause the Li abundances of RC stars to be larger than the calculation of models. In any case, the Li-rich giants with A(Li) $\ge$ 1.5 remain a puzzle. A few trigger mechanisms are proposed to explain the Li enhancement in the RC stars, such as the internal gravity waves during helium flash \citep{2020Schwab}, an additional energy loss related to the neutrino magnetic moment \citep{2021Mori}, tidal interaction in binary \citep{2019Casey}, and the merger of a helium-core white dwarf with a RGB star \citep{2020Zhangxf}.

The spectroscopy reflects the information of stellar atmosphere, which could determine the evolutionary stages as well. Based on the APOGEE Data Release 12 (DR12) data, \citet{2015Masseron} showed that a difference of [C/N] between RGB stars and helium-core burning stars can reach 0.2\,dex. It is thought to be caused by the extra-mixing along with the RGB and the helium flash. \citet{2018Hawkins} suggested that the difference ensures the spectra alone to distinguish the RGB stars from the RC stars by $Cannon$, which is a data-driven method to map the relation of spectral flux and the asteroseismic parameters. \citet{2018Ting} expanded the idea to the LAMOST DR3 low-resolution spectra with neural network method, and achieved a very low contamination rate ($\sim 3\%$) for the RC stars.

A largest sample of Li-rich giants was constructed by \citet[][\mesaone]{2019Gao} based on the low-resolution spectra from the LAMOST DR7\footnote{\url{http://dr7.lamost.org/}}. In this work, we statistically investigated their properties in detail, such as the Li-rich distribution and frequency, stellar mass, and the abundances of C and N. Our data and method are described in Section \ref{sec:sec2}. In Section \ref{sec:sec3}, we compared the individual properties between Li-rich and normal giants with the same evolutionary stages. Possible origins of Li-rich were discussed in Section \ref{sec:sec4} and a brief summary is presented in Section \ref{sec:sec5}.

\section{Data and Method} \label{sec:sec2}

LAMOST equipped 16 spectrographs with a spectral resolution (R) of about 1800. Its wavelength range is from 3800 to 9000 \AA\ \citep{2012Cui, 2012Zhao, 2022Yan}. From 2011 October to 2019 June, LAMOST DR7 has publicly released about ten of millions low-resolution spectra with SNR $\ge$ 10. The red giant stars have been selected with log $g$ $\le$ 3.5 dex and $T_\mathrm{eff}$ $<$ 5600 K, which includes the Li-rich giants from \mesaone. The typical errors for log $g$, $T_\mathrm{eff}$ and [Fe/H] are about  0.05\,dex, 30\,K and 0.03\,dex for the spectra with SNR $\ge$ 50, respectively.

\subsection{Method}  \label{sec:se21}

Data-driven methods have been applied on the LAMOST low-resolution spectra by many researchers for deriving multiple information including the asteroseismic parameters, mass/age, and elemental abundances, etc. For example, $The$ $Cannon$ \citep{2017Ho, 2018Hawkins}, KPCA \citep{2019Wu}, SLAM \citep{2020Zhangbb,2020Zhangba}, neural network \citep{2018Ting, 2019Xiang}. Suitable asteroseismic training data are required to establish the data-driven model, and the sample of \citet{2016Vrard} meets this requirement for this study, which contains 6100 giants with reliable $\Delta P$ and $\Delta \nu$. There are 2662 common stars with high-quality LAMOST spectra (SNR $\ge$ 50) between the LAMOST DR7 giant stars and \citet{2016Vrard} catalog after cross-matching. We randomly pick 1800 stars to be the training set, and the rest are the test set for verifying the generalization of the neural network model.

In this work, a neural network with three layers is constructed to describe the mapping from the spectra to the asteroseismic parameters. We derive a highly nonlinear relation between the LAMOST normalized spectra and the asteroseismic parameters (i.e., $\Delta P$ and $\Delta \nu$). The labels of $\Delta P$ and $\Delta \nu$ can be represented as a function of the normalized flux, and they can be written as:
\begin{equation}
(\Delta P, \Delta \nu) =\omega^{j}_{i}\sigma(\omega^{k}_{j}\sigma(\omega^{l}_{k}\sigma(\omega^{m}_{l}f_{m}+b_{l})+b^{k})+b_{j})+b_{i} 
\end{equation}
where $\sigma(x) = max(x, 0)$ is the activation function of Relu. $j, k, l$ are the number of neurons of 512, 256 and 64 from the first to the last layers. The flux of normalized spectra is indicated as $f_{m}$, and $\omega$ and $b$ are the coefficients to be optimized during the training process. The training process stops when the $L1$ losses of validation set achieve constant tendency after tens of thousands steps. The constant $L1$ losses indicate that the sum of difference between predicted and authentic labels ($\Delta P$, $\Delta \nu$) reaches the minimum. The training process to get a predictive model is implemented with the python $Tensorflow$\footnote{\url{https://www.tensorflow.org}} package. Following \citet{2018Ting}, the results were constraint within a convex hull of the LAMOST DR7 stellar atmospheric parameters (i.e., $T_\mathrm{eff}$, log $g$ and [Fe/H]) to exclude the extrapolation outside the range of stellar parameters of training sample.

The information of stellar mass and abundances of C and N are crucial to understanding the Li production. A similar data-driven method has been applied to derive these properties for the red giant stars. The precisions of [C/Fe] and [N/Fe] are 0.052\,dex and 0.082\,dex for stars with spectral SNR $>$ 50 \citep{2022Wang}, which would make the uncertainty of [C/N] to be about 0.1\,dex. The stellar masses with uncertainty of 0.21\,$M_{\odot}$ are estimated by mapping the relations from the high-fidelity asteroseismic mass of \citet{2018Yu} to the LAMOST spectra. The "ground truth" labels of C and N are from APOGEE DR16 with the corresponding flags = 0 \citep{2020Jonsson}.

\subsection{Data and Sample}  \label{sec:sec22}

\begin{figure*}
 \centering
   \includegraphics[width=16cm]{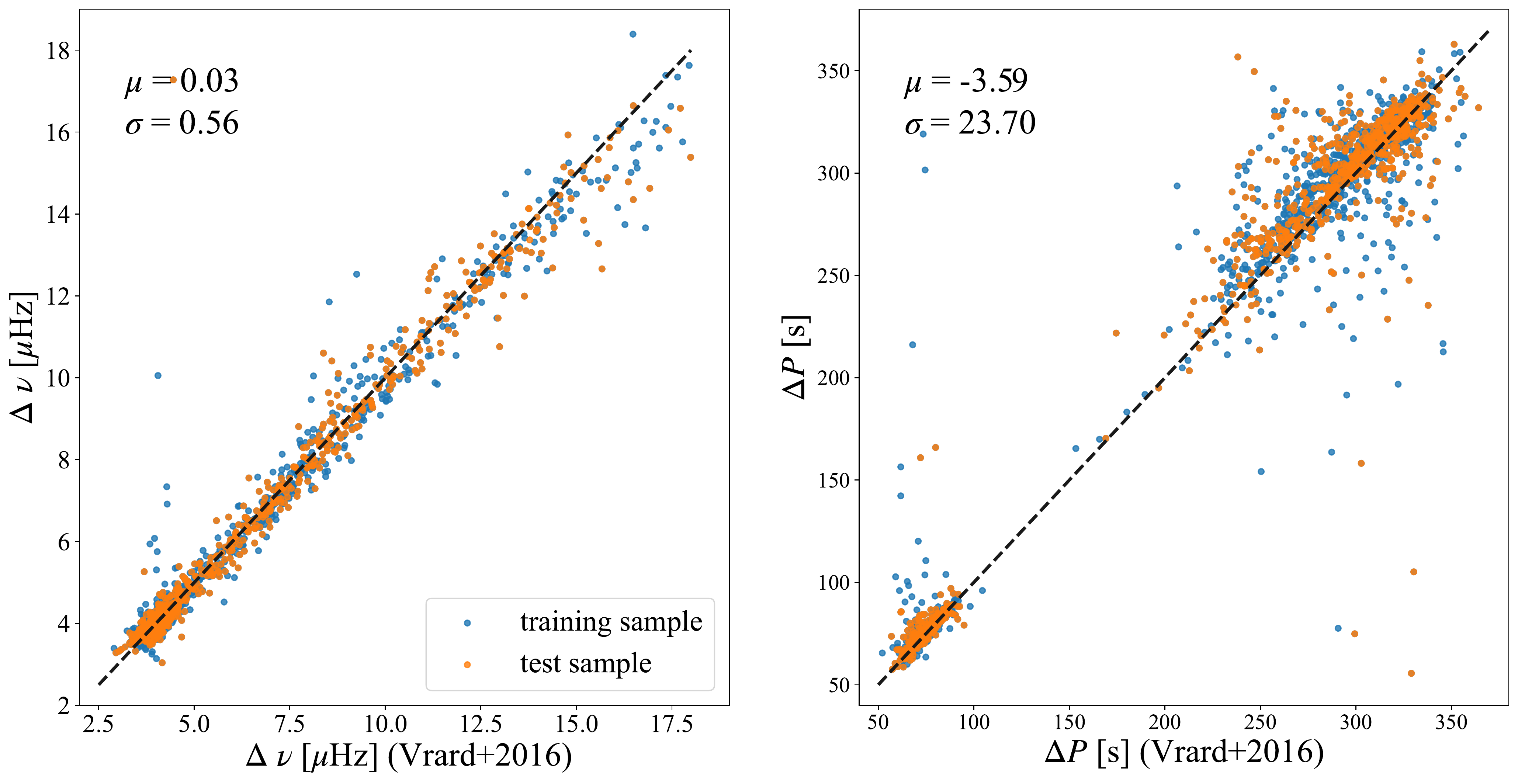}
      \caption{The comparison of $\Delta \nu$ and $\Delta P$ between the prediction from our data-driven and the values from \citet{2016Vrard}. The training sample and test sample are denoted as blue and yellow points. For both of these samples, the data-driven $\Delta \nu$ and $\Delta P$ have a good consistency with \citet{2016Vrard}, though a few outliers are shown in the $\Delta P$.}
   \label{fig1}
\end{figure*}

\begin{figure*}
 \centering
   \includegraphics[width=14cm]{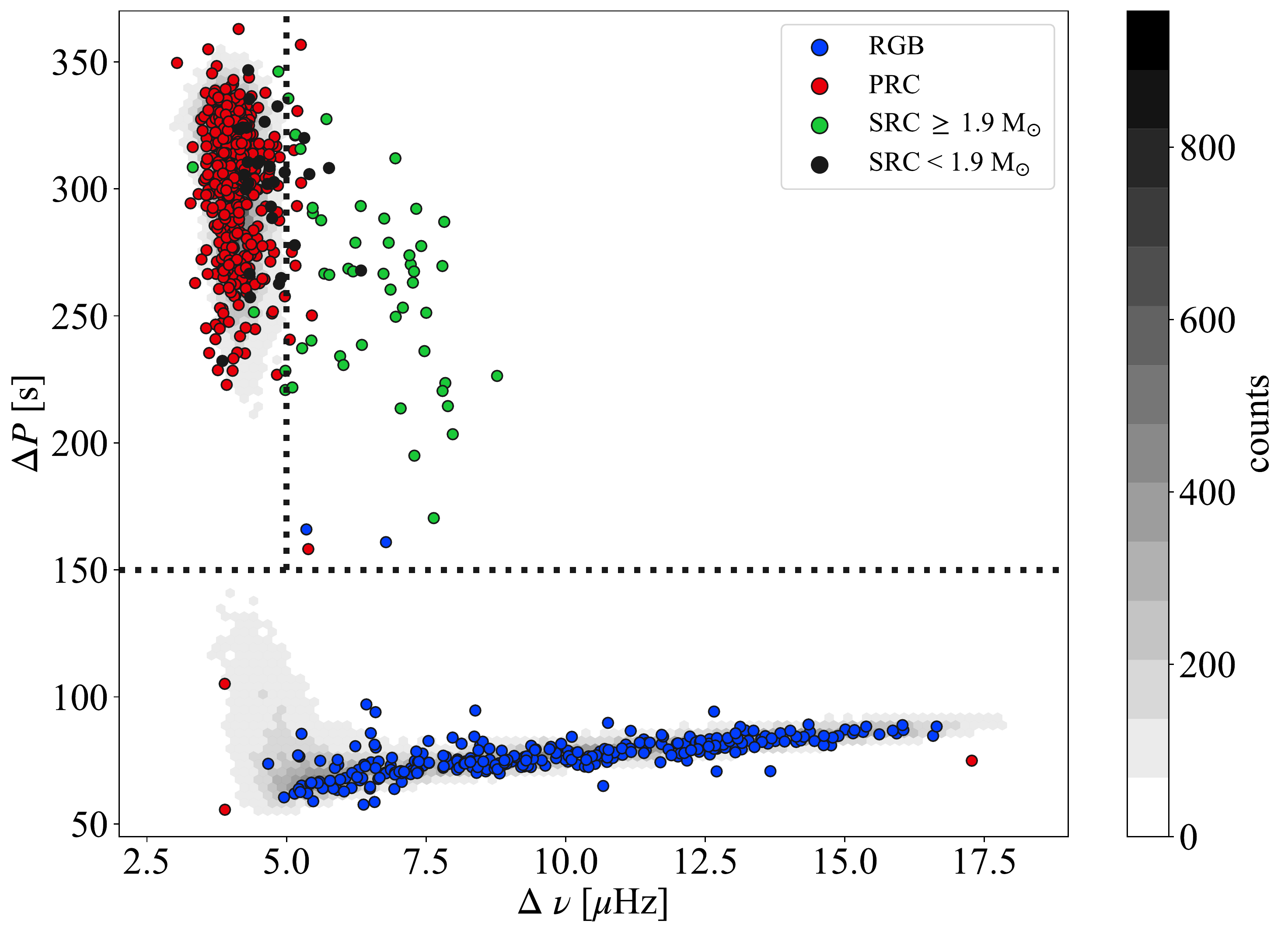}
      \caption{The diagram of $\Delta P$ versus $\Delta \nu$ for classification of stellar evolutionary stages by the data-driven method. The RGB stars and RC stars are discriminated by the horizontal line with $\Delta P$ = 150\,s, and the vertical line indicates $\Delta \nu$ = 5\,$\mu$Hz that separates the PRC stars from the SRC stars. Our final sample of giants are shown with the colour code according to the number of stars. According to \citet{2016Vrard}, the RGB, PRC, and SRC stars of test sample are presented with blue, red, and green points, respectively. The black dots are the SRC stars with a data-driven mass lower than 1.9\,$M_{\odot}$.}
   \label{fig2}
\end{figure*}

The comparison of predicted $\Delta P$ and $\Delta \nu$ and those from \citet{2016Vrard} imply an appropriate data-driven mapping from the LAMOST spectra to asteroseismic labels. Both $\Delta P$ and $\Delta \nu$ show a good consistency with one-by-one comparison, though a slightly scatter present at high $\Delta P$ (Fig.~\ref{fig1}). A clear gap can be seen between the RC and RGB stars in $\Delta P$. The RC stars occupy the high $\Delta P$ region, while the RGB stars cluster at a lower range of $\Delta P$, which are clearly separated in the diagram of $\Delta P$ versus $\Delta \nu$ (Fig.~\ref{fig2}). The criterion of $\Delta P$ = 150\,s is adopted to discriminate the RC from RGB stars suggested by \citet{2011Bedding}.

Moreover, the primary RC (PRC) stars (the low-mass stars that experience helium flash with a degenerate helium-core) could be properly identified from the secondary RC (SRC) stars ( the massive stars that ignite helium with partly or non-degenerate core) with $\Delta \nu$ $>$ 5\,$\mu$Hz \citep{2014Mosser}. We noted that part of SRC stars with low mass conflict with the fact that the SRC stars should be with a large mass. The discrepancy might be attributed to the limited number of SRC stars of the training set. For a reliable discussion, the SRC stars with mass $\ge$ 1.9\,$M_{\odot}$ have been adopted following the definition of \citet{2014Mosser} for solar metallicity PRC stars (see Section~\ref{sec:sec33}).

The test set is used to determine the completeness and purity, which is independent of the training processes of the data-driven model. Among the test set, 282, 501, and 79 stars are classified as RGB, PRC, and SRC by \citet{2016Vrard}. The model can recall 280 RGB, 485 PRC, and 49 SRC stars, so the corresponding completenesses are 99.2$\%$ (280/282), 96.8$\%$ (485/501), 62$\%$ (49/79), respectively. The model predicted 283 RGB, 515 PRC, and 62 SRC stars for the test set. Given the same number of stars at the authentic evolutionary stages, its accuracy of prediction for the RGB, PRC, SRC are 98.9$\%$ (280/283), 94.2$\%$ (485/515), 79$\%$ (49/62), respectively. The purity of SRC stars will reach 97.8$\%$ (44/45) but with a reduction of completeness down to 59.5$\%$ (44/74) if we excluded the SRC stars with masses $<$ 1.9\,$M_{\odot}$. The completeness and purity are comparable to the results of \citet{2018Ting}, who applied a similar approach for LAMOST DR3.

Finally, we get a sample of 260,396 red giant stars (SNR $\ge$ 50) with the determined evolutionary stages, including 50.9$\%$ (132,581) RGB, 43.8$\%$ (113,941) PRC, and 5.3$\%$ (12,874) SRC stars. With the method and selection criteria, 3,750 Li-rich giants have been derived with the properties of stellar mass, [C/Fe] and [N/Fe], which will be shown in Paper II (Yan et al. 2022, in preparation). The Li-rich sample consisting of 29.1$\%$ (1,093) RGB, 55.4$\%$ (2,079) PRC, and 15.4$\%$ (578) SRC stars is listed in Table~\ref{tab:tab1}.

\begin{deluxetable*}{cccccc}
\tablenum{1}
\tablecaption{Statistical distributions in the different evolutionary stages\label{tab:tab1}}
\tablewidth{0pt}
\tablehead{
\colhead{Stages} & \colhead{Total} & \colhead{Li-rich}  & \colhead{Super Li-rich} &
\colhead{}{Completeness} & \colhead{Purity}
}
\startdata
Whole sample & 260,396 (100$\%$) & 3,750 (1.4$\%$)  & 521 (13.9$\%$)  & -   & -  \\
RGB & 132,581 (50.9$\%$) & 1,093 (0.82$\%$)   & 39  (3.56$\%$)   & 99.2$\%$ & 98.9$\%$ \\
RC  & 127,815 (49.1$\%$) & 2,657 (2.08$\%$)   & 482 (18.14$\%$)  & 92.1$\%$ & 92.5$\%$ \\
PRC & 113,941 (43.8$\%$) & 2,079 (1.82$\%$)   & 471 (22.66$\%$)  & 96.8$\%$ & 94.2$\%$ \\
SRC & 12,874  (5.3$\%$)  & 578   (4.17$\%$)   & 11  (1.9$\%$)    & 59.5$\%$ & 97.8$\%$ \\
\enddata
\tablecomments{The stars are divided into the evolutionary stages of RGB, RC, PRC and SRC. For the column of Total, the percentage means the fraction of the stars at each stage. For the columns of Li-rich and Super Li-rich, the percentages represent the probabilities of the normal stars to be Li-rich and the probabilities of the Li-rich stars to be super Li-rich (i.e. Li-rich/Total, Super Li-rich/Li-rich). Here, we adopted the stars with A(Li) $\ge$ 1.5 as Li-rich and stars with A(Li) $\ge$ 3.3 as super Li-rich.}
\end{deluxetable*}

\section{Results} \label{sec:sec3}
\subsection{Distribution of Li abundances} \label{sec:sec31}

\begin{figure}
 \centering
  \includegraphics[width=8.8cm]{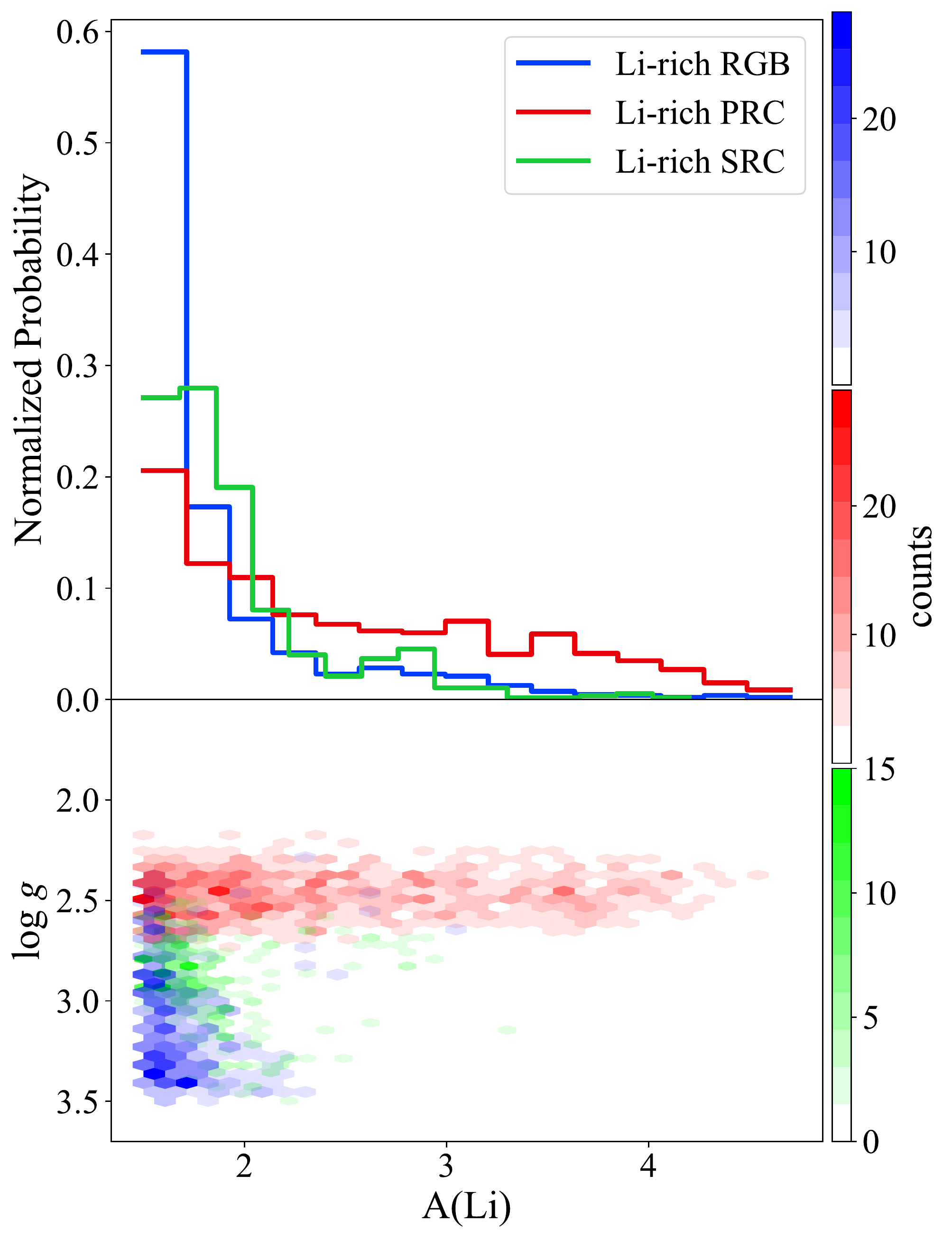}
      \caption{The distributions of Li abundances for different evolutionary stages. In the top panel, the Li-rich giants in the RGB, PRC, and SRC are denoted by blue, red, and green, respectively, which each bin of Li abundances is about 0.20\,dex in the histogram. Note that they have been normalized by the number of Li-rich stars at the corresponding evolutionary stage. The bottom panel is the Li abundance versus log $g$ with colors in number density.} 
  \label{fig3}
\end{figure}

The Li-rich stars are rare in the red giant stage. Previous studies have found that about 1$\%$ $\sim$ 2$\%$ giants show Li excess (i.e., A(Li) $\ge$ 1.5, \citealt{2019Gao, 2019Casey, 2021Martell}). A similar ratio (1.4$\%$) of Li-rich giants is found for our giants' sample. 

A large fraction of the Li-rich giants have been confirmed as the RC with slightly different quantities, which are 68$\%$ \citep{2019Deepak}, 80$\%$ \citep{2019Casey}, 86$\%$ \citep{2021Yan}, and 64$\%$ \citep{2021Martell}, respectively. There are 70.9$\%$ of our Li-rich giants belong to the RC. The different ratios might be attributed to the selection function, classified method, spectral resolution, and the determination of Li abundance. 

It is found that the percentage of Li-rich at the RC stage is 2.08$\%$, which is 2.54 times of that (0.82$\%$) of RGB stars (Table~\ref{tab:tab1}). Similar probabilities of RC and RGB stars to be Li-rich have been found by \citet{2021Martell} (1.9$\%$; 0.76$\%$) based on the GALAH and K2-HERMES surveys.

\citet{2021Yan} found that the distribution of Li-rich RC stars significantly differ from those of the Li-rich RGB stars. Fig.~\ref{fig3} presents the distributions of Li abundances at the three evolutionary stages for our sample. The number of Li-rich stars decreases as the Li abundance increases. It is generally believed that less stars have a higher A(Li) because it is hard to maintain the Li due to further depletion. 

For the Li-rich RGB stars, Li distribution more rapidly decreases compared to the Li-rich RC stars, which is in line with \citet{2021Yan}. They showed a decreasing number towards the low log $g$, and more Li-rich RGB stars are found at higher log $g$ as the bottom panel of Fig.~\ref{fig3} shown. This trend was also found by \citet{2021Zhangjh} in the view of A(Li) versus stellar radius, which involves the RGB stars with A(Li) = 0 $\sim$ 2.5 based on the data of LAMOST medium-resolution and $Kepler$. A few exceptions (3.56$\%$) of Li-rich RGB stars were shown to be with super high Li, which might be contaminated by the RC stars. It requires a firm verification by the direct asteroseismic data.

For the two subgroups of RC stars, the Li-rich ratio on the SRC (4.17$\%$) is higher than that of PRC stars (1.82$\%$). Conversely, the probability (22.66$\%$) of Li-rich PRC stars to be super Li-rich is much higher than that of the Li-rich SRC stars (1.9$\%$). As found by \citet{2021Martell}, although the Li-rich ratio of PRC stars is lower than that of SRC stars, the PRC stars are more likely to be super Li-rich. As the bottom panel of Fig.~\ref{fig3} shown, it is obvious that a few SRC stars exhibit very high A(Li) similar to the RGB stars. 

In contrast to the Li-rich SRC stars, the Li-rich PRC stars spread at a large Li abundance range within a narrow log $g$, though the number decreases toward the high A(Li) as well. Since the ranges of metallicity are wider for the PRC stars relative to the SRC ones \citep{1999Girardi}, \citet{2021Martell} suggested that it would lead to more complexity of Li related processes within the PRC stars.


\subsection{Li-rich frequency} \label{sec:sec32}
\begin{figure}
 \centering
  \includegraphics[width=8.5cm]{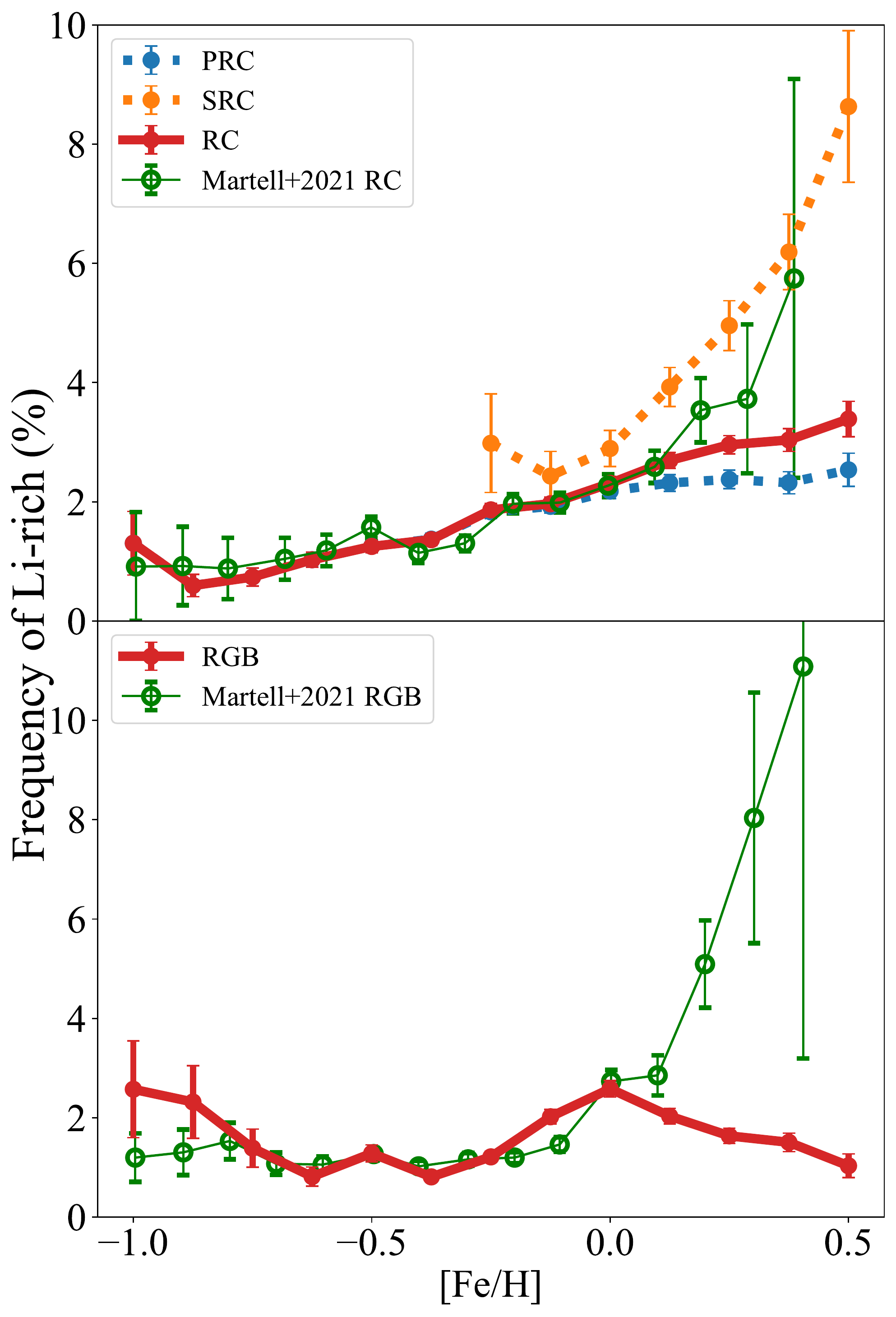}
      \caption{The frequency of Li-rich giants versus metallicity among the different evolutionary stages. The Li-rich frequency at each evolutionary stage is estimated for a metallicity bin of $\sim$ 0.1\,dex. The frequencies of Li-rich stars from \citet{2021Martell} are also presented as green points for comparison. Top panel: the frequency of Li-rich RC (red), Li-rich PRC (blue), and SRC stars (orange). Bottom panel: the frequency of Li-rich for RGB stars (red).}
  \label{fig4}
\end{figure}

The frequency of Li-rich stars among the field \citep{2014Adamow} and the open clusters \citep{2016Smiljanic} have been identified to be 1 $\sim 2\%$. The Li-rich frequency likely increase towards high metallicity \citep{2019Casey, 2020Deepak, 2021Martell}. The different mechanisms of Li enhancement are usually tied to the stellar evolution with different metallicity, it is worth exploring the Li-rich fraction not only in a given evolutionary stage but also in the metallicity bins.

Fig.~\ref{fig4} shows that the Li-rich RC stars occur more frequently with increasing of [Fe/H], especially for the Li-rich SRC stars. Interestingly, the frequency of Li-rich SRC stars agrees with the prediction ($3.4 \sim 9.6\%$) of the high-mass RGB stars by \citet{2016Aguilera}, and they originally attempt to explain the Li-rich RGB stars by engulfment of substellar companions. The Li-rich frequency among the massive RGB stars ($\ge$ 1.9\,$M_{\odot}$) was estimated to be a high percentage of about 8.5$\%$ similar to the SRC stars.

The frequency of RC stars to be the Li-rich gradually grows toward higher [Fe/H], although it is not so sharp as \citet{2021Martell}. It can be seen that the pattern is more consistent with that of \citet{2021Martell} if we combine the groups of PRC and SRC stars. The minor difference above the solar metallicity is negligible within the error bars. Moreover, other than dramatically increase as \citet{2021Martell}, the frequency of Li-rich RGB stars presents a turning point near the solar metallicity. At the low metallicity end, the Li-rich frequency increases for both RC and RGB stars, which need further confirmation due to the limited number of metal-poor stars.

In the super-solar metallicity, our Li-rich frequencies are lower than that of \citet{2021Martell} for the RGB stage. The disagreement might be due to the selection function and the classified method. On one hand, it is noted that two samples have different fractions of Li-rich RGB. We got a lower fraction (29$\%$) of Li-rich RGB stars than 42$\%$ of \citet{2021Martell}. Fig.~\ref{fig4} shows that our error bars are relatively small at the super-solar metallicity, which implies a more reliable statistical analysis. The error bar is adopted to be the Poisson statistics error same as \citet{2021Martell}. On the other hand, our classifications of evolutionary stages achieve both higher purity and completeness. For instance, our RGB stars are determined with 98.9$\%$ purity and 99.2$\%$ completeness, while \citet{2021Martell} derived the RGB sample with 81$\%$ purity and 74$\%$ completeness, which would be more contaminated by the RC stars.

\subsection{Mass distribution} \label{sec:sec33}
\begin{figure}
 \centering
   \includegraphics[width=8.5cm]{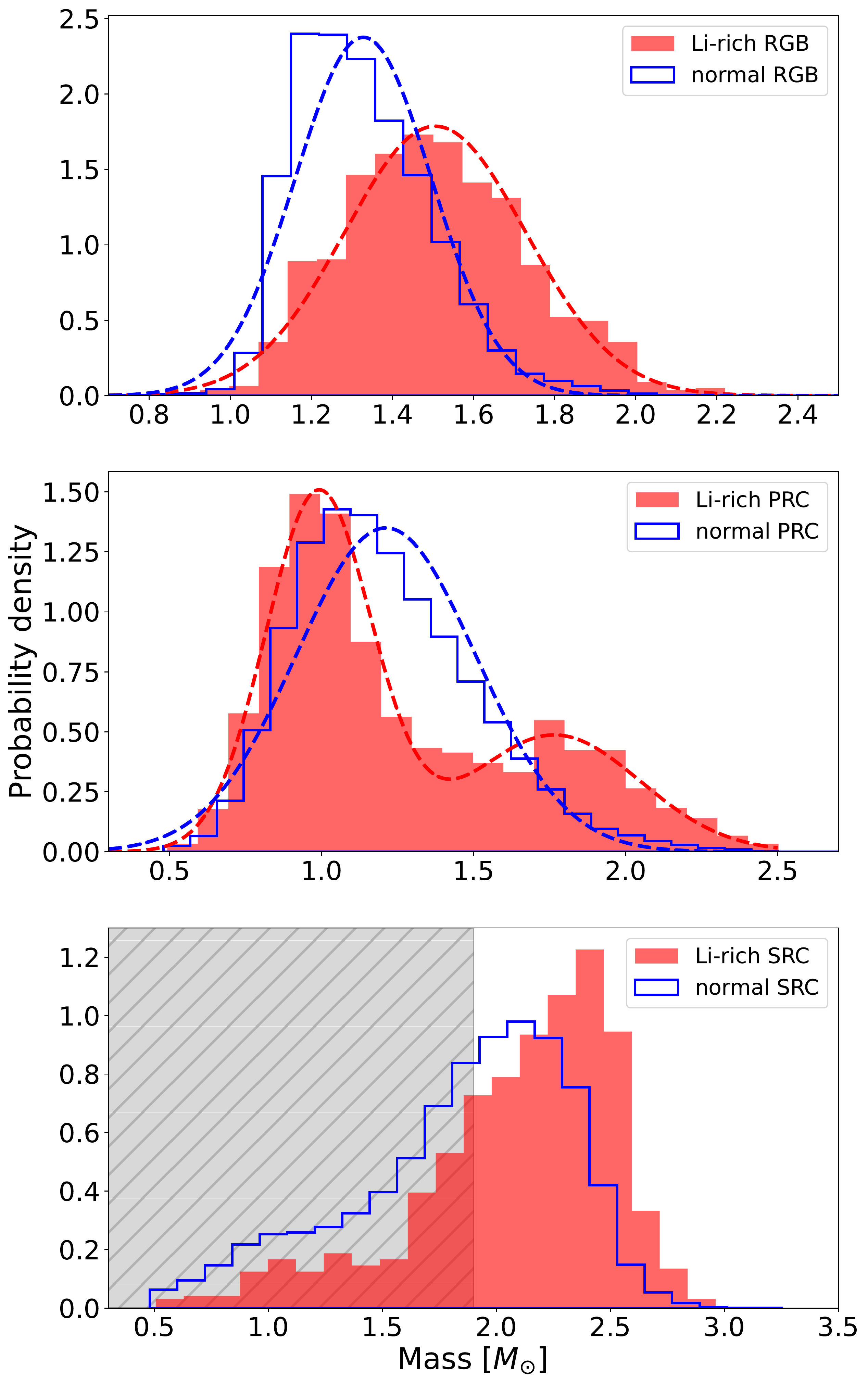}
      \caption{The distributions of stellar masses for the RGB, PRC, and SRC stars (from top to bottom panel). The Li-rich giants are indicated as the red histogram, while the blue for normal giants. The dashed lines are the fitting results with GMMs.} In the bottom panel, the grey hatched regions indicate stars with mass less than 1.9\,$M_{\odot}$, which are not included in the discussion.
   \label{fig5}
\end{figure}

The extra mixing mostly depends on the stellar mass, metallicity, and evolutionary phase, which may impact the Li enhancement and dilution \citep{2010Charbonnel, 2019Shetrone, 2021Chaname}. Therefore, it is important to analyze the mass distribution of our Li-rich giants. 

\citet{2020Deepak} found that the mass distribution of Li-rich stars at/below RGB Bump show to be similar as the normal stars. With a more extensive Li-rich sample, our data show that the Li-rich RGB stars concentrate at a higher mass of 1.51\,$M_{\odot}$ than the Li-normal RGB stars of 1.33\,$M_{\odot}$ assuming a gaussian profile, which is in agreement with the asteroseismic mass distribution of \citet{2021Yan} for RGB stars. 

Rather than a straightforward mass definition on the SRC stars, the PRC sample could not be simply confined by a single mass definition considering their large metallicity range. The mass criteria to separate the PRC from SRC would vary from 1.7 to 2.5\,$M_{\odot}$ \citep{2000Castellani, 2016Girardi}, depending on the stellar models with different input physical parameters (e.g., convective mixing of the MS, initial composition and metallicity).

With respect to the normal PRC stars, the Li-rich PRC stars present a bimodal distribution, which can be fitted with the Gaussian mixture models (GMMs). The normal stars assemble at the mass of 1.21\,$M_{\odot}$, while the Li-rich PRC stars seem to cluster at the mass of 1.0\,$M_{\odot}$ and 1.77\,$M_{\odot}$ (Fig.~\ref{fig5}). Even though the PRC stars might be slightly contaminated by the SRC stars, they still cover the intermediate-mass range of 1.5 $\sim$ 2.0\,$M_{\odot}$. It is in accord with \citet{2021Yan} and \citet{2021Deepak} for the RC stars, but noted that both of their RC sample include some of SRC stars, which have not been picked out from their RC sample.

For the pre-RGB stars in the open clusters, a bimodality is presented in the diagram of stellar masses versus Li abundances. In other words, there is a gap in the mass range of about 1.2 $\sim$ 1.4\,$M_{\odot}$, which is known as the ``Li dip" \citep{1986Boesgaard, 2019Deliyannis, 2020Twarog}. It implies a relatively high Li abundances in the pre-RGB stars outside the region of Li dip. These stars would be highly possible to become the Li-rich when they climb towards the giant branch. It is found that a bimodal mass distribution is presented for the more evolved Li-rich PRC stars, but not for the Li-rich RGB ones as shown in Fig.~\ref{fig5}. A further discussion will be presented in Section~\ref{sec:sec4}.

For the SRC stars, if we only considered the SRC stars with moderately high mass above 1.9\,$M_{\odot}$ suggested as \citet{2014Mosser}, the Li-rich and normal SRC stars do not show obviously different mass distributions, but with a slightly higher mass for the Li-rich stars. The distribution of SRC stars can not be well fitted with a gaussian function due to the narrow mass range. As we can see from Fig.~\ref{fig5}, part (26$\%$) of the Li-rich SRC stars locate at the low-mass region, which is possible due to the contamination (21$\%$) from others. However, a relatively large ratio of the normal SRC stars with low mass is probably caused by a limited number of RC with high mass to train the data-driven model. Other than the PRC stars, the SRC stars only cover a narrow range of metallicity. To ensure a reliable discussion for the SRC stars, the SRC sample with mass greater than 1.9\,$M_{\odot}$ is adopted for the statistical analysis.

\subsection{Carbon and Nitrogen} \label{sec:sec34}

\begin{figure*}
 \centering
   \includegraphics[width=17cm]{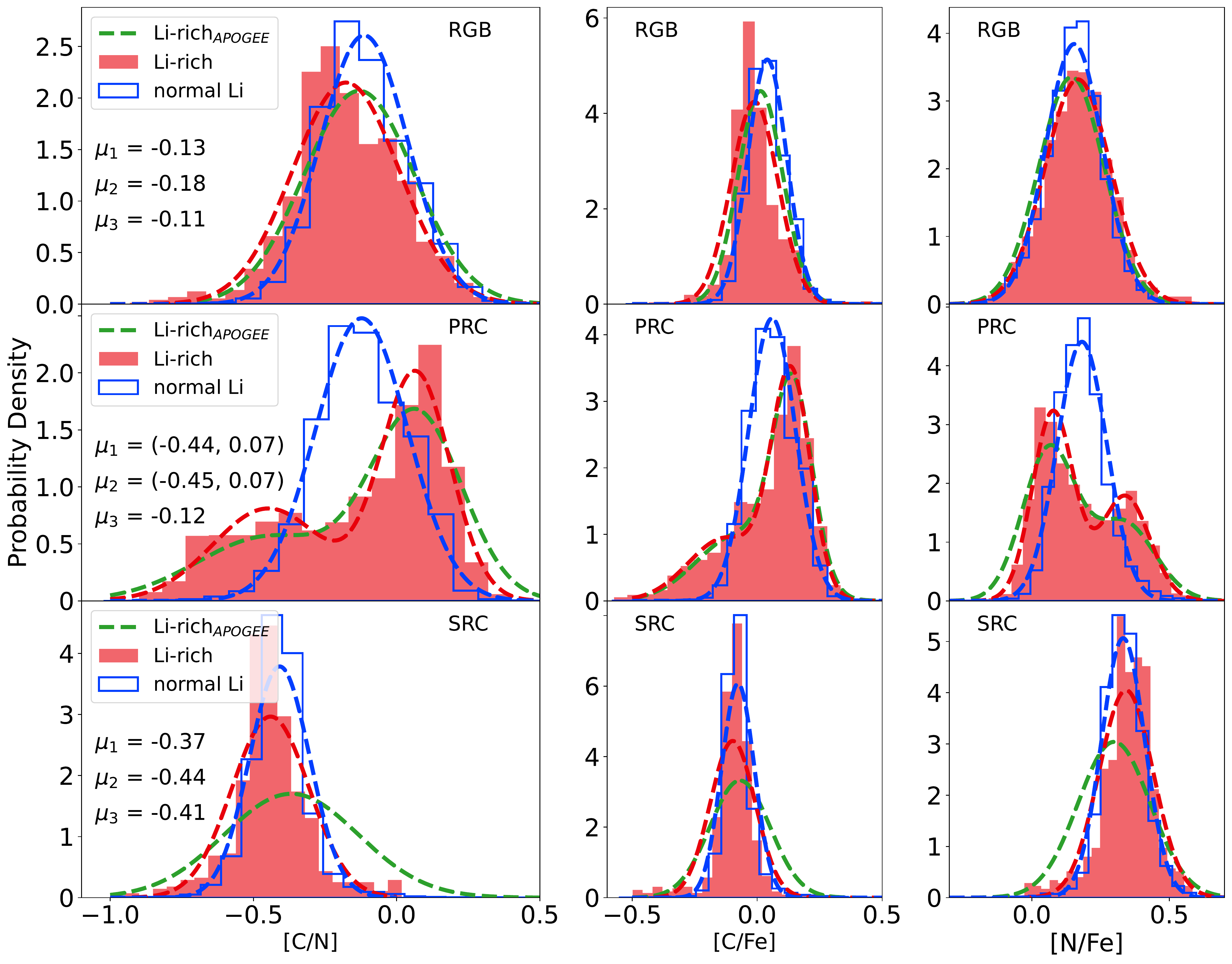}
      \caption{The probability distribution of [C/N], [C/Fe], and [N/Fe] for the giants. [C/N], [C/Fe] and, [N/Fe] are displayed at the columns from right to left. Three rows from top to bottom correspond to stages of RGB, PRC, and SRC. The Li-rich and normal giant stars are shown as red and blue. The green indicates the subsample from APOGEE DR16. $\mu$ with subscripts from 1 to 3 represent the centers of [C/N] profiles of the Li-rich APOGEE, Li-rich and normal Li stars.}
   \label{fig6}
\end{figure*}                    

\begin{figure*}
 \centering
   \includegraphics[width=17cm]{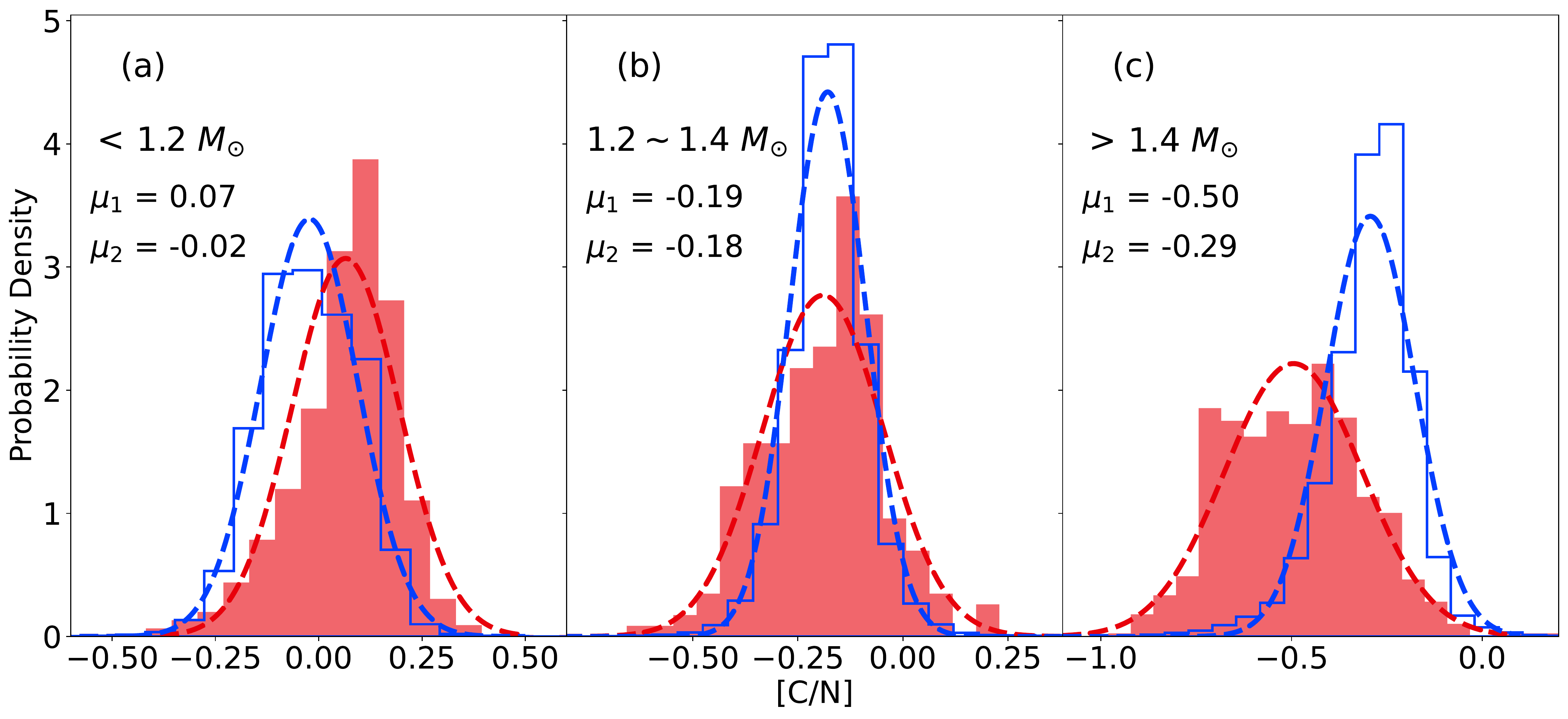}
      \caption{The [C/N] distribution of the three subsamples of PRC stars. The panels of a,b,c indicate the PRC stars with masses below, near or above the Li dip, respectively. The color indications are same as Fig.~\ref{fig6}. For each panel, the centers of [C/N] distributions of Li-rich and normal PRC stars are denoted as $\mu_{1}$ and $\mu_{2}$.}
   \label{fig7}
\end{figure*}

The abundances of carbon and nitrogen are valuable to trace the extra mixing processes. The FDU will transport the CNO-cycle processed materials with deficient C and enhanced N from the hot interior to the surface, causing the decline of [C/N] at the outer layer \citep{1967Ibenb}. After the FDU, a further decrease of [C/N] would be taken place caused by the extra mixing at the RGB bump \citep{2010Charbonnel, 2017Masseron, 2019Lagarde, 2019Shetrone}. The ratio of [C/N] is then sensitive to the internal mixing that might be responsible for the Li-rich.

In this context, the comparisons of [C/N] between the Li-rich and normal giants for different evolutionary stages are shown in Fig.~\ref{fig6}. To confirm the comparison, a subsample of Li-rich giants with C and N from APOGEE DR16 is shown as green, which contains 138 RGB, 287 PRC, and 78 SRC stars. GMMs are applied to model the distributions of [C/Fe], [N/Fe] and [C/N] for our sample stars.

For the Li-rich RGB stars, the distribution of [C/Fe] looks like those of normal stars only with minor offset. Their [N/Fe] almost share a same distribution with normal RGB stars in Fig.~\ref{fig6}. The [C/N] ratios cluster at similar values of -0.18 and -0.11 for the Li-rich and normal RGB stars, respectively.

Similarly, in the bottom panels of Fig.~\ref{fig6}, the Li-rich SRC stars are indistinguishable from the normal giants in the [C/N] distribution, although [C/N] of the Li-rich SRC stars from the APOGEE data are not perfectly consistent with our sample due to the small number. It indicates that most of the Li-rich RGB and SRC stars might not experience any internal Li-enriched process, since [C/N] would exhibit different distribution between Li-rich and normal RGB/SRC stars if an unusual mixing process operates to produce Li.

Compared to the other evolutionary phases, the Li-rich PRC stars display a bimodal [C/N] distribution as their mass distribution, differing from the normal PRC stars with only one peak. Regardless of the whole sample or the subset from APOGEE, two gaussian functions are good enough to interpret the distributions especially for [N/Fe] and [C/N]. The Li-rich PRC stars mainly center on the [C/N] with -0.44 and 0.07\,dex, while [C/N] of the normal PRC stars locate on a medium value of -0.12. The [C/Fe] seems distribute with two peaks but are not prominent as [N/Fe]. The feature of [N/Fe] is consistent with that of \citet{2021Yan}.

The bimodality occurs in both of mass and [C/N] for the Li-rich PRC stars, which is suspect to be the reflection of Li dip shown in the MS stars of open clusters. Analogous to the study of Li in open clusters, we divide the PRC sample into three groups by the stellar masses below, near and above the Li dip (i.e., $<$\,1.2$M_{\odot}$, \,1.2$M_{\odot} \sim 1.4\,M_{\odot}$, and $>$\,1.4$M_{\odot}$). As Fig.~\ref{fig7} shown, the Li-rich PRC stars below, near and above the Li dip cluster at the high, medium and low [C/N], respectively. It is clear to see that the Li-rich stars below and above the Li dip correspond to the two peaks in Fig.~\ref{fig6}. However, these two major portions exhibit contrary [C/N] behaviours compared to their Li-normal counterparts. The Li-rich PRC stars with mass $<$ 1.2$\,M_{\odot}$ have larger [C/N] than those of normal ones, but [C/N] of the Li-rich above Li dip shift towards a lower value relative to the normal stars. For the mass region of Li dip, [C/N] of the Li-rich and normal PRC stars behave with a similar center of about -0.19. It could be speculated that the distribution of Li-rich PRC stars might be affected by Li dip, and the Li-enriched processes of PRC stars would be associated with the stellar mass and result in the different [C/N] behaviour.

\section{Discussion} \label{sec:sec4}
\subsection{Origins of the Li-rich on RGB}  \label{sec:sec41}

Recently, only minor fraction of Li-rich stars are identified as the RGB stars with the help of asteroseismology \citep{2021Yan, 2021Singh}. The information of pre-RGB can provide most important constraints on the understanding of Li-rich RGB.

Firstly, \citet{2021Yan} found that the distribution of Li-rich RGB stars dramatically decline with an increase of Li abundance. With extensive sample size, we found that the distribution of the Li-rich RGB stars is anti-correlated with Li abundance but positively related to log $g$. It indicates that as the stars evolve more towards low log $g$, less RGB stars show to be the Li-rich due to the depletion. We do not find any uniform Li-enriched process at the specific evolutionary stage of RGB (e.g. RGB bump) similar to \citet{2021Zhangjh}.

Secondly, the Li-rich frequency of RGB stars against metallicity might provide another clue. It looks like connect to the Li abundances of MS field stars. For the MS stars, the maximum Li abundance, A(Li)$_\mathrm{max}$, climbs up to the solar metallicity and drop at [Fe/H] $>$ 0 \citep{2018Bensby, 2018Fu}, which agrees with the pattern of Li-rich RGB frequency. Different dominated Li origins (e.g. novae, AGB stars) for the Galactic Li would result in such pattern \citep{2019Cescutti, 2019Guiglion}. As the ``initial" A(Li) varies with metallicty, the solar metallicity MS stars with higher A(Li)$_\mathrm{max}$ would be more likely to be Li-rich. Thus, the probability of Li-rich RGB stars is expected to be associated with the metallicity.

Thirdly, for the MS stars of the open clusters, the distribution of stellar masses present two peaks with high Li abundances, i.e., the morphology of Li dip. These MS stars with high A(Li) is most likely to be the Li-rich RGB stars, thereby it is supposed to present the bimodal morphology of mass for the Li-rich RGB. However, it is not found in our Li-rich sample, which only show higher masses than the Li normal RGB stars.

The speculation is that some Li-enriched processes probably have happened, for example the engulfment of the planet or substellar companion. Recently, the calculation of \citet{2021Soares} showed that the planetary engulfment would be more efficient to enrich Li for the post MS stars with 1.4$\,M_{\odot}$. Their models suggest that Li-enriched stars would be pile-up among 1.4$\sim$1.6\,$M_{\odot}$, which is consistent with mass distribution of our Li-rich RGB stars. Note that these scenarios are limited to account for the Li-rich stars with meteoritic A(Li), but a small part of RGB stars show to be the super Li-rich (i.e., A(Li) $>$ 3.3), which we could not exclude the possibility of binary interaction suggested by \citep{2016Casey}.

Additionally, the similar [C/N] between Li-rich and normal RGB stars agrees with the suggestion that the Li-rich RGB stars unlikely go through the internal mixing process to produce Li. Considering stellar properties of Li-rich RGB stars associated the information from the pre-RGB, the possible origins for the Li-rich on RGB stars are the consequence of Li depletion from the Li-rich progenitors and/or caused by the engulfment of planet or substellar companions.

\subsection{Li-rich or Li-normal SRC stars?}  \label{sec:sec42}

The Li-rich RC stars are usually correlated with the helium flash since all the low-mass RC stars are supposed to go through this violent process. While, the Li-rich phenomenon on the SRC can not be connected with the scenario of helium flash because the massive SRC stars evolve too fast to develop a degenerate helium core \citep{1999Girardi}. Therefore, another explanation is required to account for the Li-rich phenomenon on SRC.

Similar to the Li-rich RGB stars, few of the Li-rich SRC stars present super high Li abundances, and they have a consistency of Li-rich frequency with the massive RGB stars. According to the PAdova and TRieste Stellar Evolution Code \citep[PARSEC,][]{2012Bressan}, a star with 2\,$M_{\odot}$ only spends about 14 Myr on the RGB phase, it is too short to allow the star to experience any significant mixing at RGB. Thus, when those stars evolve to SRC, they are mostly supposed to inherit the Li abundances of high-mass RGB stars and the corresponding Li-rich frequency.

If it is true, the high-mass RGB stars should be with higher Li abundance than the SRC stars, because they would endure further Li depletion as they are more evolved. Given only Li abundances provided for the Li-rich giants, a rough comparison show that the mean A(Li) of massive Li-rich RGB stars ($\ge$ 1.9\,$M_{\odot}$) is higher with 0.22 dex than that of Li-rich SRC stars. The Li-rich phenomenon on SRC is probably the natural dilution from the Li-rich RGB stars with high mass.

There is only a slight difference on the mass distribution between Li-rich and normal SRC stars. Although the Li depletion has a mass dependency, it becomes less sensitive as the mass increases towards the higher range \citep{2021Deepak}. It fits in with the fact that the high-mass stars have a shallow convective envelope to make the Li depletion less effective. Moreover, the information of [C/N] indicates that the extra mixing process would not happen for producing fresh Li in the massive SRC stars. As \citet{2014Adamow} reported, the massive Li-rich stars show similar stellar properties with normal giants and no evidence points out the specific Li-rich process.

The Li-rich SRC stars might be the natural consequence of Li depletion from the Li-rich high-mass progenitors on the RGB. The probability to be Li-rich on SRC might depend on not only the Li depletion at SRC but also the initial Li at the massive RGB.

\subsection{The Li enhancement of PRC stars} \label{sec:sec43}

\begin{figure}
 \centering
   \includegraphics[width=8cm]{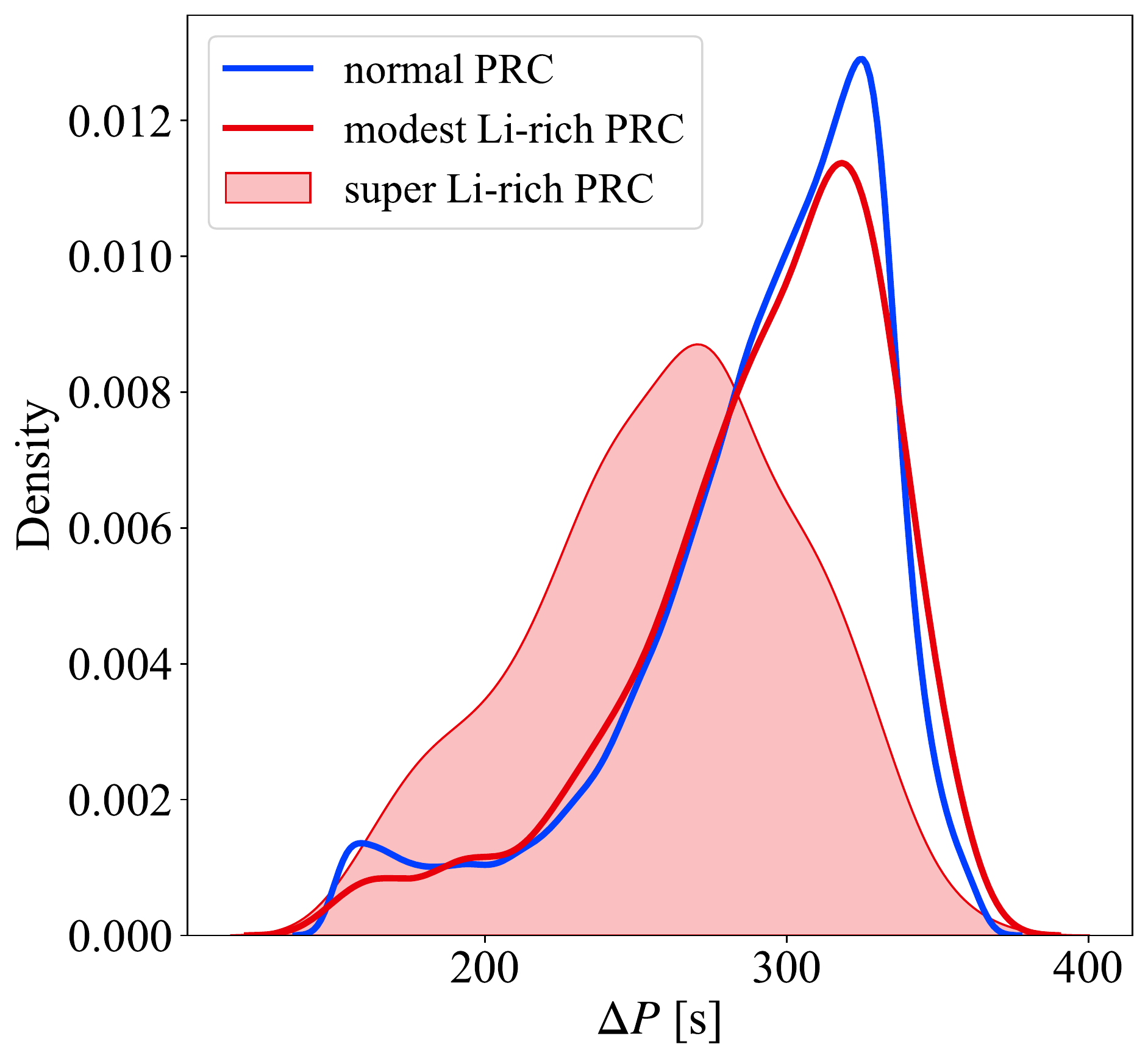}
      \caption{The density distribution of the asymptotic gravity-mode period spacing $\Delta P$ for the normal Li ( A(Li) $<$ 1.5), modest Li-rich with 1.5 $\le$ A(Li) $<$ 3.3 and super Li-rich of PRC stars.}
   \label{fig8}
\end{figure}

Since the majority of the Li-rich giants are identified as the RC stars rather than the RGB stars by the asteroseismology, the origins to enrich Li have attracted the community attention to how Li to be produced at this phase \citep{2021Yan, 2021Singh}. 

Based on the GALAH DR2, \citet{2020Kumar} even reported that all of PRC stars have undergone the Li-enriched process due to the helium flash, because they found an increase of 40 factors in Li abundance from the RGB tip to the RC for the low mass stars, where the RC stars concentrate around A(Li) = 0.7 contradicted the prediction of A(Li) $\sim$ -0.9 by the stellar models. So they suggested a ubiquitous Li enhancement on the RGB tip or the helium core ignition. However, \citet{2021Chaname} claimed the increase of observed Li can be well reproduced by their models if considering different Li depletion rates with various stellar masses. They argued that the mass adopted by \citet{2020Kumar} is higher than the asteroseismic mass, thus it leads to a high Li at the RC due to the lower Li dilution for higher mass. Although it could explain the modest Li ($\sim$ 1.0\,dex) during the stage of PRC, the PRC stars with A(Li) $\ge$ 1.5 are indeed required another scenario.

To solve the discrepancy between the standard stellar models and observation of \citet{2020Kumar}, the novel physical processes, such as neutrino magnetic moment \citep{2021Mori} and internal gravity waves \citep{2020Schwab}, have been invoked to trigger the mixing to produce Li during the RGB tip or helium flash. But neither of the predictions of these models could achieve super Li-rich. 

By comparing the predictions of stellar models and the observed asteroseismic parameters, a novel insight is proposed by \citet{2021Singh} to study the status of Li-rich during the RC stage. They reported that the super Li-rich PRC stars are exclusively located at the early stage of core-helium burning with low $\Delta P$ = 257\,s and the modest Li-rich stars are intermediate stage with $\Delta P$ = 290\,s between the early super Li-rich and the late Li-normal ($\Delta P$ = 306\,s). With their speculation, all super Li-rich should assemble at the early stage of RC. However, by inspecting the stellar evolution with the view of mass and radius, \citet{2021Zhangjh} found that the super Li-rich stars scatter at the RC phase rather than cluster at the early stage. Following \citet{2021Singh}, we displayed the $\Delta P$ density distribution of PRC stars as the indicator of evolution (Fig.~\ref{fig8}). With a large number of super Li-rich PRC stars, we find that the super Li-rich truly concentrate at the early stage with $\Delta P$ = 260\,s, which is consistent with \citet{2021Singh}. But they span a similar range of $\Delta P$ as Li-normal and modest Li-rich rather than a clear distinction as shown by \citet{2021Singh}. It suggests that the Li-rich processes might happen during the early stage of core-helium burning.

Rather than the Li-rich process triggered by internal mechanisms, \citet{2004Denissenkov} suggested the rapid rotation to enhance the extra mixing in the binary is responsible for the Li-rich stars at the RGB bump. This scenario of binary interaction is generalized for both the RGB and RC by \citet{2019Casey}. Based on the data of a nine-year radial-velocity monitoring, \citet{2020Jorissen} found a similar binary frequency between Li-rich and normal giants, therefore, the Li enhancement is not supported by the tidal interaction between binary systems.

We cross-match the Li-rich stars with the catalog of close-binary systems provided by \citet{2020Price} based on multiple observations from APOGEE DR16. We can not find a higher binary fraction of the Li-rich giants (4.3$\%$, 4/92) than that of normal giants (6.4$\%$, 350/5471). Additionally, \citet{2020Zhangxf} proposed that the Li-rich RC stars are formed as the post-merger of a RGB stars and a helium white dwarf. Although it could reproduce the mass distribution of our PRC stars with this model of near solar metallicity, it does not account for the [C/N] and the change of Li-rich frequency with metallicity. The model needs to be confined with more information (e.g. [C/N]) due to the poor knowledge of the common envelope of a merger as \citet{2020Zhangxf} stated.

As mentioned in Section~\ref{sec:sec33}, the MS stars outside mass range of Li dip have so high A(Li) that they are very likely to be the Li-rich. The bimodal distribution is not presented for the Li-rich RGB stars because of the planetary engulfment on the post MS. But it is shown at the more evolved Li-rich PRC stars, which is likely related to the Li dip (see Section~\ref{sec:sec34}). Even though the stars outside the Li dip have high ``initial" Li abundance so that impact the distribution of Li-rich PRC stars, the processes of Li enhancement should be existed to be responsible for Li-rich PRC due to the further Li depletion during the RGB. In addition, when we divide the PRC stars into the below, near and above the mass regions of Li dip, [C/N] still exhibit different distributions between the Li-rich and normal PRC stars, which might be the reflection of the Li-enriched mixing processes.

\section{Summary} \label{sec:sec5}

A large sample of giants from LAMOST DR7 were classified into the evolutionary stages of RGB, PRC, and SRC by a data-driven approach. The purities can reach 97.9$\%$ for RGB stars, 93.6$\%$ for PRC stars, and 76.6$\%$ for SRC stars with the LAMOST spectral SNR $\ge$ 50. We established the sample of red giant stars with the informations of mass, elemental abundances of C and N, aiming to interpret the Li-rich problem at different evolutionary stages by the statistical analysis of the Li-rich distribution and properties between Li-rich and normal giants. 

We find that most of the Li-rich RGB stars might be the descendant of Li-rich pre-RGB stars, and the scenario of engulfment of planet or substellar companions are responsible for their Li-rich. The Li-rich PRC stars, which dominate among the Li-rich giants, are probably interpreted by internal mixing processes near the helium flash. According to their [C/N], the Li-rich process seems depend on the stellar mass. Whereas, the Li-rich phenomena for the massive SRC stars are the consequence of natural Li depletion.

Although several Li-rich models have been proposed to explain the Li-rich PRC stars, more information is required to be predicted to match observed features, such as Li-rich frequency, stellar mass, and [C/N]. From the view of observation, only a small fraction of giant stars are derived with the ratio of $^{12}$C/$^{13}$C, which is more sensitive to the mixing process. In this work, similar to the other data-driven approaches, our study is limited by the stellar parameters of the training set in order to exclude the extrapolation.

\vspace{7mm} \noindent {\bf Acknowledgments}

We thank the anonymous referee for the valuable comment to improve the paper. Y.-T.Z. thanks Dr.~Sanjib Sharma for discussion about the stellar mass. This work was funded by the National Key R$\&$D Program of China No. 2019YFA0405500 and the National Natural Science Foundation of China (NSFC) under grant No.11973001, 12090040, 12090044, 12022304, 11973052, 11973042, 11833006, and U1931102. Y.-T.Z. and B.Z. acknowledges support from the LAMOST FELLOWSHIP fund. H.-L.Y. acknowledges support from the Youth Innovation Promotion Association of the CAS (id. 2019060), and NAOC Nebula Talents Program. YST acknowledges financial support from the Australian Research Council through DECRA Fellowship DE220101520.

The LAMOST FELLOWSHIP is supported by Special Funding for Advanced Users, budgeted and administered by the Center for Astronomical Mega-science, Chinese Academy of Sciences (CAMS-CAS). This work is supported by the Cultivation Project for LAMOST Scientific Payoff and Research Achievement of CAMS-CAS.

Guoshoujing Telescope (the Large Sky Area Multi-Object Fiber Spectroscopic Telescope LAMOST) is a National Major Scientific Project built by the Chinese Academy of Sciences. Funding for the project has been provided by the National Development and Reform Commission. LAMOST is operated and managed by the National Astronomical Observatories, Chinese Academy of Sciences.

$Facility:$ LAMOST.

$Software:$ Astropy \citep{2013Astropy, 2018Astropy}, Matplotlib \citep{2007Hunter}, TOPCAT \citep{2005Taylor}.

\bibliography{lirichdr7}{}
\bibliographystyle{aasjournal}

\end{document}